\documentclass[aps,prd,twocolumn,superscriptaddress,showpacs,floatfix]{revtex4-1} 
\usepackage{graphicx}
\usepackage{dcolumn}   
\usepackage{bm}        
\usepackage{amssymb}   
\usepackage{color}
\usepackage{ifluatex}
\usepackage[utf8]{\ifluatex lua\fi inputenc}
\usepackage[T1]{fontenc}
\usepackage{newtxtext,newtxmath}
\usepackage{epstopdf}
\usepackage{slashed}
\usepackage{lineno}

\usepackage[FIGTOPCAP]{subfigure}
\usepackage{stackengine}

\begin{document}
\title{Hyperon production in $\Lambda_c^+\to K^-p\pi^+$ and $\Lambda_c^+\to K^0_Sp\pi^0 $}
\author{Jung Keun Ahn}
\affiliation{Department of Physics, Korea University, Seoul 02841, Republic of Korea}
\author{Seongbae Yang}
\affiliation{Institute for Basic Science, Korea University, Seoul 02841, Republic of Korea}
\author{Seung-il Nam\footnote{E-mail: {\tt sinam@pknu.ac.kr}}}
\affiliation{Department of Physics, Pukyong National University (PKNU), Busan 48513, Republic of Korea}
\affiliation{Asia Pacific Center for Theoretical Physics (APCTP), Pohang 37673, 
Republic of Korea}\date{\today}

\date{\today}

\begin{abstract}
We investigate $S=-1$ hyperon production from 
the $\Lambda_c^+\to K^-p\pi^+$ and $\Lambda_c^+\to K^0_Sp\pi^0$ decays
within the effective Lagrangian approach. 
We consider the $\Sigma/\Lambda$ ground states, $\Lambda(1520)$, 
$\Lambda(1670)(J^p=1/2^-)$,
$\Lambda(1890)(J^p=3/2^+)$; $\Lambda/\Sigma$-pole contributions from the 
combined resonances between 1800 MeV and 2100 MeV; and $N/\Delta$-pole
and $K^\ast$-pole contributions, which include the proton, $\Delta(1232)$,
and $K(892)$.
We calculate the Dalitz plot density
$(d^2\Gamma/dM_{K^-p}dM_{K^-\pi^+}$) for the $\Lambda_c^+\to K^-p\pi^+$
decay. The calculated result is 
in good agreement with experimental data from the Belle Collaboration. 
Using the parameters from the fit, we present 
the Dalitz plot density for the $\Lambda_c^+\to K^0_Sp\pi^0$ decay. 
In our calculation, a sharp peak-like structure
near 1665 MeV is predicted in the $\Lambda_c^+\to K^-p\pi^+$
decay because of the interference effects between  
the $\Lambda(1670)$ resonance and $\eta$-$\Lambda$ loop channels.
We also demonstrate that 
we can access direct information 
regarding the weak couplings of $\Lambda(1670)$ and $\Sigma(1670)$
from the $\Lambda_c^+\to K^0_Sp\pi^0$ decay. 
Finally, a possible interpretation for the 1665 MeV structure beyond
our prediction is briefly discussed.
\end{abstract}
\maketitle


\section{Introduction}

In the constituent quark model, low-lying baryons with 
$J^p=1/2^+$ and $J^p=3/2^+$ make up the
ground-state 56-plet in approximate flavor-spin SU(6) multiplets. 
Odd-parity baryons are classified into a band with orbital excitation 
$L=1$, which entails $P=-1$; in combination with $S=1/2$ 
or $3/2$, this gives negative-parity baryons 
with $J^p=1/2^-$, $3/2^-$, and $5/2^-$. However, 
the excited states of hyperons are still much less well known 
compared with the nucleon resonances. 
Thus, studying hyperon resonances may provide some hints regarding 
the role of confinement in the nonperturbative QCD region.

In the $S=-1$ sector, only a few states are directly measured 
in production experiments, whereas other broad states are studied 
in multichannel particle-wave analyses, mostly 
with $\overline{K}N$ scattering data. 
For example, only the $\Lambda(1520)(J^p=3/2^-)$ 
above the $\overline{K}N$ threshold is reconstructed from 
its decay channels, $\pi\Sigma$, $\overline{K}N$, and $\pi\pi\Lambda$. 
Other $\Lambda^\ast$ and $\Sigma^\ast$ resonances are overlaid 
with relatively large decay widths so that it is challenging 
to identify their lineshapes separately from the others in the invariant
mass spectra. 

In the mass region from 1600 to 2000 MeV, 8 $\Lambda^\ast$ and 
5 $\Sigma^\ast$ resonances are listed in the Particle Data Group (PDG)
tables with three- and four-star ratings \cite{Tanabashi:2019oca}. 
$\Lambda(1600)(J^p=1/2^+)$ and $\Sigma(1660)(J^p=1/2^+)$ lie below 
the $\eta\Lambda$ threshold (1663.5 MeV) and 
are known to have a strong coupling to the $\overline{K}N$, 
$\pi\Lambda$ and $\pi\Sigma$ channels \cite{Gao:2012}. 
The next $\Lambda(1670)(J^p=1/2^-)$ and $\Sigma(1670)(J^p=3/2^-)$ are 
very close to the $\eta\Lambda$ threshold. 
The $\Lambda(1670)$ is interpreted as a resonance strongly coupled 
to a pure $I=0$ $\eta\Lambda$ state. 
The nature of the $\Sigma(1670)$ is still poorly known, and 
the production angular distribution of 
the $K^-p\to\Sigma(1670)^+\pi^-$ reaction is interpreted as evidence for
two mass-degenerate $\Sigma(1670)$ resonances \cite{Eberhard:1969}. 
One couples strongly to $\pi\Sigma$, and the other couples 
to $\pi\pi\Sigma$.   
$\Lambda(1690)(J^p=3/2^-)$ decays largely to $\pi\Sigma(1385)$. 
Above 1700 MeV,   
another eight $\Lambda^\ast$ and $\Sigma^\ast$ resonances with three- 
and four-star ratings appear near each other in the mass range 
up to 2000 MeV.   

A recent observation of hidden-charm pentaquark states 
reported by the LHCb Collaboration emphasizes the importance of
understanding $\Lambda^\ast$ and $\Sigma^\ast$ resonances 
in the $K^-p$ invariant mass spectrum for $\Lambda_b\to J/\psi K^-p$ 
decays \cite{LHCb1:2015, LHCb2:2019}. 
In the charm sector, possible evidence for a new $\Lambda^\ast$
resonance at a mass of approximately 1665 MeV, 
just above $\eta\Lambda$ threshold, 
has been reported from the Belle Collaboration in the $K^-p$ invariant 
mass spectrum for $\Lambda_c^+\to K^-p\pi^+$ decays \cite{Tanida:2018}. 
The new $\Lambda^\ast$ resonance shows a narrow peak 
with a Breit-Wigner width of approximately 10 MeV, 
which could be interpreted 
as either a dynamically generated $\Lambda(1671)(J^p=3/2^+)$ 
\cite{Kamano:2015hxa}, $\Lambda(1667)(J^p=3/2^-)$ in the $D_{03}$ 
partial wave \cite{BChao:2012}, or an exotic $\Lambda^\ast$ state. 
More recently, the peak-like structure was interpreted 
using the threshold cusp, 
enhanced by the triangle singularities \cite{Liu}. 

In the $\Lambda^+_c\to\overline{K}N\pi$, 
an isospin $I=0$ amplitude of the $\overline{K}N$ system dominates 
compared with the $I=1$ amplitude 
because the $\Lambda_c^+$ is an iso-singlet state and 
the transition amplitude $c\to su\overline{d}$ has 
$\Delta I=1$ with $I=1$ pion emission \cite{Miyahara:2015cja}. 
Therefore, excited $\Lambda$ hyperons can be 
selectively produced in the $\overline{K}N$ invariant mass spectrum.
Conversely, the large branching fraction of 
$\Gamma(\Lambda^+_c\to\Lambda\pi^0\pi^+)/\Gamma_{\rm total}=(7.1\pm 0.4)$\% 
\cite{Tanabashi:2019oca, Ablikim:2017} supports 
a possible population of excited $\Sigma$ hyperons
decaying to $\overline{K}N$, as $\Lambda\pi^0$ is a pure $I=1$ state. 

Therefore, the $\Lambda^+_c\to\overline{K}N\pi$ decays are good probes 
to test the isospin symmetry in non-leptonic decays of the charmed baryon. 
The $I=1$ $\Sigma^\ast$ resonances can only be involved in 
the $\Lambda_c^+\to K^0_Sp\pi^0$ decays, 
while the $I=0$ $\Lambda^\ast$ resonances are dominant in 
the $\Lambda_c^+\to K^-p\pi^+$ decays. 
In this respect, it is necessary to conduct measurements of
the $\Lambda_c^+\to K^0_Sp\pi^0$ decay. A charged $K^0_Sp$ system
ensures a production of $\Sigma^{\ast+}$ hyperons 
isolated from $\Lambda^\ast$ hyperons, 
thereby providing a good opportunity to test isospin symmetry. 

Moreover, a possible interference
effect among $\overline{K}N$, $\overline{K}\pi$, 
and $\pi N$ channels is also very interesting. 
A strong $K^\ast$ band crossing the $\Lambda(1520)$ band shows evidence for
interference between $K^\ast$ and $\Lambda(1520)$ production channels in the
$\Lambda_c^+\to K^-p\pi^+$ decay \cite{Yang:2015ytm}. 
The phase in the interference between the two resonances 
can be deduced from experimental data. 
   
In this paper, we report numerical calculation results for $S=-1$ 
hyperon production from the $\Lambda_c^+\to K^-p\pi^+$ and 
$\Lambda_c^+\to K^0_Sp\pi^0$ decays within
the effective Lagrangian approach. 
We consider the $\Sigma/\Lambda$ ground states, $\Lambda(1520)$, 
$\Lambda(1670)(J^p=1/2^-)$,
$\Lambda(1690)(J^p=3/2^-)$; $\Lambda/\Sigma$-pole contributions from the 
combined resonances between 1800 MeV and 2100 MeV; and $N/\Delta$-pole
and $K^\ast$-pole contributions, including the proton, $\Delta(1232)$,
and $K(892)$. 

We calculate the Dalitz plot density
$(d^2\Gamma/dM_{K^-p}dM_{K^-\pi^+}$) for the $\Lambda_c^+\to K^-p\pi^+$
decay, which is in good agreement with experimental data 
from the Belle Collaboration.
Using the coupling constants from the fit, we present 
the Dalitz plot density for the $\Lambda_c^+\to K^0_Sp\pi^0$ decay. 
In our calculation, a sharp resonance-like structure
near 1665 MeV is predicted to appear
because of the interference effect between 
$\Lambda(1670)$ production and $\eta$-$\Lambda$ channels.
We also demonstrate that 
we can access direct information regarding weak couplings 
of $\Lambda(1670)$ and $\Sigma(1670)$ 
from the $\Lambda_c^+\to K^0_Sp\pi^0$ decay. 
Finally, a possible interpretation for the 1665 MeV structure beyond
our prediction is briefly discussed.

\section{Theoretical framework}

\begin{figure}[tbph!]
\includegraphics[width=4.1cm]{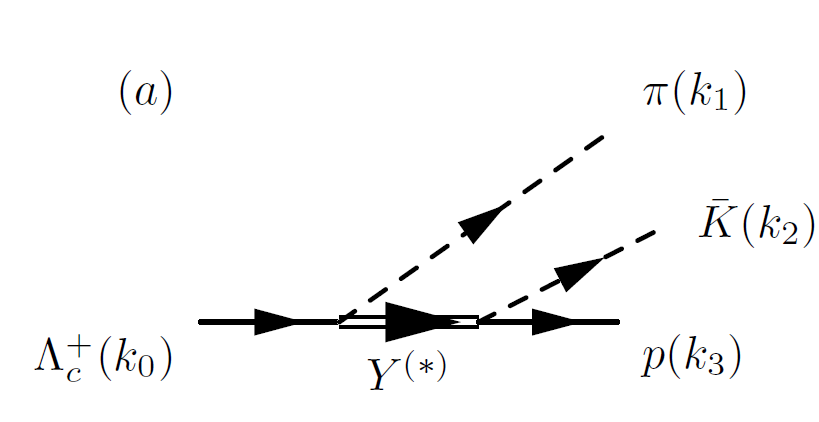}
\includegraphics[width=4.1cm]{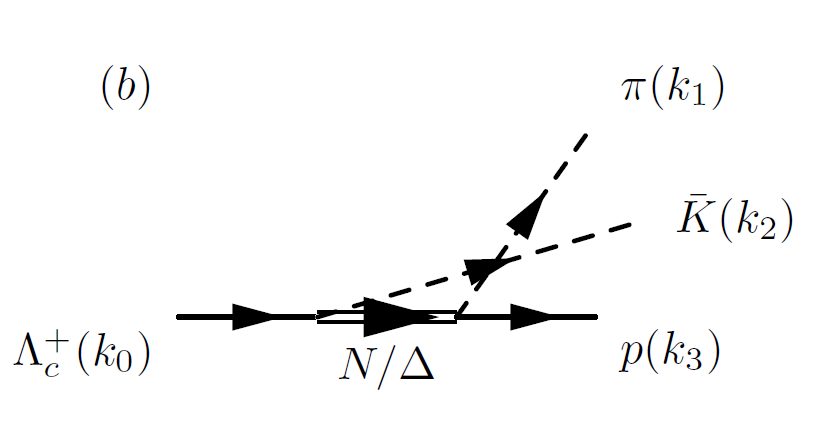}
\includegraphics[width=4.1cm]{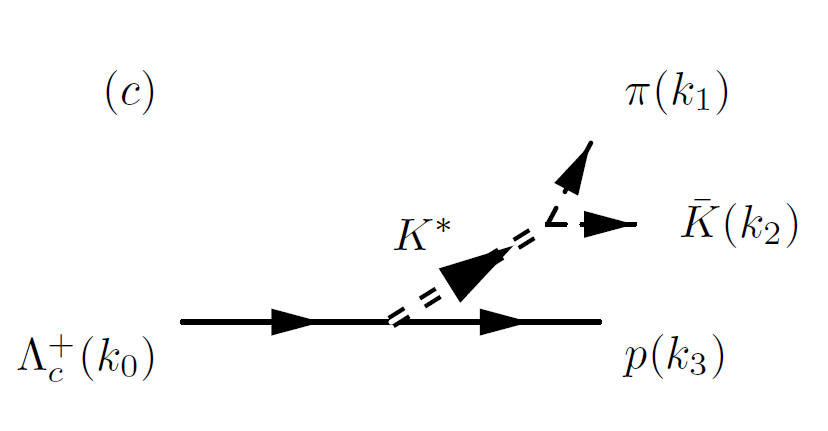}
\includegraphics[width=4.1cm]{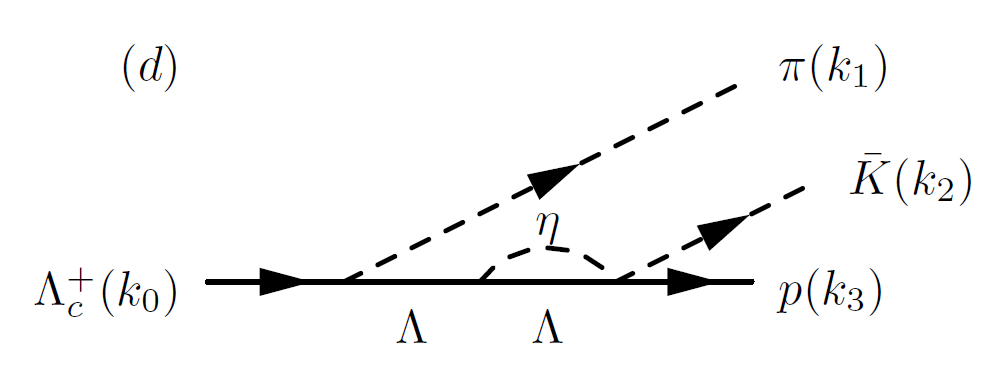}
\caption{Relevant Feynman diagrams for $\Lambda^+_c\to\pi\overline{K}p$: 
($a$) $Y^{ (*)}$-pole diagram, ($b$) $N/\Delta$-pole diagram, ($c$) $K^*$-pole diagram, and 
($d$) $\eta$-$\Lambda$ loop diagram,
where $Y=(\Lambda,\Sigma)$. See the text for details.}       
\label{fig:fig1}
\end{figure}

In this Section, we introduce the theoretical framework to study 
the hadronic $\Lambda^+_c$ decay within the effective Lagrangian approach.
We consider the \textit{charged} $(\Lambda^+_c\to\pi^+K^-p)$ 
and \textit{neutral} $(\Lambda^+_c\to\pi^0\overline{K}^0p)$ decay channels. 
Relevant Feynman diagrams for the two channels are illustrated 
in Fig. ~\ref{fig:fig1}.
The diagrams in Fig.~\ref{fig:fig1}($a$), $(b)$, ($c$), ($d)$ denote 
$Y^{(*) }$-pole, $N/\Delta$-pole, 
$K^*$-pole, and $\eta$-$\Lambda$-loop diagrams, respectively. 
Although tens of baryon resonances can be accessible from the $\Lambda^+_c$ 
decay~\cite{Tanabashi:2019oca}, we take into account only a few resonances
to minimize theoretical uncertainties and control numerical calculations. 

Selection criteria for the resonances in the numerical calculations 
are based on the following points. 
First, by investigating the Dalitz plot of 
the Belle experimental data for the charged channel given in Fig.~3 of 
Ref.~\cite{Yang:2015ytm} (and in Fig~\ref{fig:fig1}(a) 
in the present work), we observe clear signals from $\Lambda(1520)$, 
$\Lambda(1670)$ (or $\Sigma(1670)$), $\Delta(1232)$, and $K(892)$. 
Second, there are possible non-resonant backgrounds (BKG) shown 
in the lower $M^2_{K^-p}$ region, 
which can be explained by the ground-state $\Lambda$ and $\Sigma$. 
Third, excluding $\Delta(1232)$, 
whose contribution provides a diagonal band in the Dalitz plot, 
there are no obvious non-strange resonances. 
This observation can be understood as the color factors of 
the $u$ and $\overline{d}$ quarks from $W^+$ decay are
constrained to form the color-singlet pion and resonances. 
Hence, we only take into account the ground-state nucleon and 
$\Delta(1232)$.
Finally, the authors of Ref.~\cite{Miyahara:2015cja} suggested that, 
even in the Cabibbo-favored (CF) decays such as those in the diagram ($a$), 
the isospin $I=1$ hyperon-resonance $(\Sigma^*)$ contributions 
are suppressed because of the strong $[ud]$ di-quark correlation inside 
$\Lambda^+_c$, whereas the $I=0$ contributions $(\Lambda^*)$ 
prevail for the $\Lambda^+_c$ weak decay with $\pi^+$.  

\begin{table*}[t]
\centering
\begin{tabular}{c|c||c|c|c|c|c|c|c|c} \hline
\multicolumn{2}{c||}{}&\multicolumn{4}{c|}{ $\Lambda^+_c\to\pi^+K^-p$}
&\multicolumn{4}{c}{ $\Lambda^+_c\to\pi^0\overline{K}^0p$}\\
\hline
$B(M,J^P)$&$\Gamma$ [MeV]&$g_{K^-pB}$&$g_{\pi^+\Lambda^+_cB}$&$g_{\pi^+p B}$
&$g_{K^-\Lambda^+_cB}$&$g_{\overline{K}^0p B}$&$g_{\pi^0\Lambda^+_c B}$
&$g_{\pi^0pB}$&$g_{\overline{K}^0\Lambda^+_cB}$\\
\hline
$p(938,1/2^+)$&$0$&$-$&$-$&$-$&$-$&$-$&$-$&$13$&$(-7.11,25.90)$\\
$\Delta(1232,3/2^+)$&$117$&$-$&$-$&$2.17$&$\underline{0.09}$&$-$&$-$&$1.77$&$\underline{0.09}$\\
$\Lambda(1116,1/2^+)$&$0$&$-13.4$&$(-4.6,15.8)$&$-$&$-$&$-$&$-$&$-$&$-$\\
$\Lambda(1520,3/2^-)$&$15.6$&$10.92$&$\underline{-0.006}$&$-$&$-$&$-$&$-$&$-$&$-$\\
$\Lambda(1670,1/2^-)$&$35$&$1.62$&$\underline{-0.11}$&$-$&$-$&$-$&$-$&$-$&$-$\\
$\Lambda(1890,3/2^+)$&$150$&$0.67$&$\underline{0.1}$&$-$&$-$&$-$&$-$&$-$&$-$\\
$\Sigma(1193,1/2^+)$&$0$&$4.09$&$(5.4,-2.7)$&$-$&$-$&$5.78$&$(5.4,-2.7)$&$-$&$-$\\ \hline
$K(892,1^-)$&$50$&\multicolumn{2}{c|}{$g_{\pi^+K^-K^{*0}}=3.76$}&
\multicolumn{2}{c||}{$g_{K^{*0}p\Lambda^+_c}=\underline{-0.77}$}&
\multicolumn{2}{c|}{$g_{\pi^0\overline{K}^0K^{*0}}=-2.66$}&
\multicolumn{2}{c}{$g_{K^{*0}p\Lambda^+_c}=\underline{-0.77}$} \\ \hline
\end{tabular}
\caption{Relevant inputs for numerical calculations: 
Full decay widths and strong and weak decay constants for the hadrons involved~\cite{Sharma:1998rd}. 
The values in parentheses represent 
the parity violating (PV) and parity conserving (PC) couplings 
as $(g^\mathrm{PV},g^\mathrm{PC})$. 
The weak couplings are in units of 
$G_FV_{ud}V^*_{cs}10^{-2}\,\mathrm{GeV}^2\approx(1.11\times10^{-7})$. 
The underlined values indicate those determined by reproducing 
the Belle experimental data for $\Lambda^+_c\to \pi^+ K^-p$~\cite{Yang:2015ytm}.}
\label{tab:tab1}
\end{table*}

Combining these observations and discussions, 
we can minimize the number of relevant contributions 
for the $\Lambda^+_c$ decay into seven and four contributions 
for the charged and neutral channels, respectively, as shown 
in Table~\ref{tab:tab1}. 
The table provides the quantum numbers, full widths, 
and relevant coupling constants. 
The underlined values in the table indicate 
those fitted to reproduce the charge-channel data.

There is one caveat: If the $I=0$ $uds$-quark cluster dominates 
the $\Lambda^+_c$ hadronic decays together with $\pi^+$ in the final state, 
as suggested in Ref.~\cite{Miyahara:2015cja}, one may expect 
$\Gamma_{\Lambda^+_c\to\pi^+\Sigma^0}/\Gamma_{\Lambda^+_c\to\pi^+\Lambda}\approx0$, for instance. 
On the contrary, this decay ratio turns out be almost unity experimentally 
~\cite{Tanabashi:2019oca}. 
Similarly, considering the isospin decompositions 
of the final state of the $\Lambda^+_c\to\pi^+\overline{K}N$ 
in the isospin limit, 
the decay ratio of the neutral and charged channels as in the present work 
becomes unity because $\mathcal{A}^{(1)}$ disappears in Eq.~(35) of 
Ref.~\cite{Lu:2016}. 
However, this is not the case~\cite{Tanabashi:2019oca},
although there can be more complicated contributions, 
such as the higher-mass hyperon and $\Delta$ resonances, 
$K^*$ contribution, and interferences. 
Hence, although we have reduced theoretical uncertainties 
using the $I=0$ meson-baryon channel dominance 
in the final state~\cite{Miyahara:2015cja}, actual experimental data 
can exhibit sizable $I=1$ $\Sigma^*$-resonance contributions that are
different from the present numerical results, 
which illustrates the future experimental data qualitatively.

Once the relevant contributions are fixed, the effective Lagrangians 
for the interaction vertices shown in Fig.~\ref{fig:fig1} are 
defined in general as follows:
\begin{eqnarray}
\label{eq:LAG}
\mathcal{L}^\mathrm{weak}_{PBB}&=&i
(\overline{B}\Gamma_5\gamma_5)P(g^\mathrm{PV}_{PBB}-g^\mathrm{PC}_{PBB}\gamma_5)B+\mathrm{h.c.},
\cr
\mathcal{L}^\mathrm{weak}_{PBB'}&=&\frac{i}{M_P}
(\overline{B}'_\mu\Gamma_5)(\partial^\mu P)(g^\mathrm{PV}_{PBB'}-g^\mathrm{PC}_{PBB'}\gamma_5)B+\mathrm{h.c.},
\cr
\mathcal{L}^\mathrm{strong}_{PBB}&=&-ig_{PBB}(\overline{B}\Gamma_5)PB+\mathrm{h.c.},
\cr
\mathcal{L}^\mathrm{strong}_{PBB'}&=&-\frac{ig_{PBB'}}{M_P}
(\overline{B}'_\mu\Gamma_5\gamma_5)(\partial^\mu P)B+\mathrm{h.c.},
\cr
\mathcal L_{VBB} &=&g_{VN \Lambda_c}
\overline{B}\slashed{V}\gamma_5(g^\mathrm{PV}_{VBB}-g^\mathrm{PC}_{VBB}\gamma_5)B+ \mathrm{h.c.},
\cr
\mathcal L_{PPV}&=&ig_{PPV}
V_\mu\left[P\partial^{\mu}P^\dagger-P^\dagger\partial^{\mu}P\right]+ \mathrm{h.c.},
\end{eqnarray}
where $P$, $B$, $B'$, and $V$ represent the $0^-$ pseudoscalar meson, $1/2^\pm$ baryon, $3/2^\pm$ baryon, 
and $1^-$ vector meson, respectively. $\Gamma_5$ defines the parity for the \textit{created} baryon 
($\overline{B}$) by
\begin{equation}
\label{eq:PARI}
\Gamma_5=\Big\{
\begin{array}{c}
\gamma_5\,\,\,\mathrm{for}\,\,\,1/2^+,\,3/2^+,\\
1_{4\times4}\,\,\,\mathrm{for}\,\,\,1/2^-,\,3/2^-.
\end{array}
\end{equation}
In principle, the PC and PV vertices 
have different coupling strengths. 
Note that Ref.~\cite{Sharma:1998rd}, 
using the heavy-quark effective field theory (HQEFT), 
provides the PC and PV weak couplings for $\Lambda^+_c$ decays 
into octet hadrons. 
For instance, $g^\mathrm{PV}_{\pi^+ \Lambda\Lambda^+_c}=-4.6$ 
and $g^\mathrm{PC}_{\pi^+ \Lambda\Lambda^+_c}=15.8$ 
in units of $G_FV_{ud}V^*_{cs}10^{-2}\mathrm{GeV}^2$. 
Here, $G_F$ and $V_{ud,cs}$ indicate the Fermi constant and 
CKM matrix elements, respectively. 
However, regarding $\Lambda^+_c$ decays into hyperon resonances, 
there is little experimental or theoretical information from which 
to extract relevant PC and PV weak couplings. 
Hence, to reduce theoretical uncertainties, we assume that 
$g^\mathrm{PC}_{PY^*\Lambda^+_c}=g^\mathrm{PV}_{PY^*\Lambda^+_c}=g_{PY^*\Lambda^+_c}$ 
for the hyperon resonances and determine 
the value of $g_{PY^*\Lambda^+_c}$ 
using the experimental data, as described previously. 

The invariant amplitudes for the diagrams in Fig.~\ref{fig:fig1} 
can be computed straightforwardly 
using the effective Lagrangians defined above. 
The total invariant amplitude is as follows:
\begin{widetext}
\begin{eqnarray}
\label{eq:INVAMP}
i\mathcal{M}^{(a)}_B&=&ig_{KNB}
\frac{\overline{u}_p\Gamma_5\Delta(q_{2+3})\Gamma_5\gamma_5
(g^\mathrm{PV}_{\pi B\Lambda_c}-g^\mathrm{PC}_{\pi B\Lambda_c}\gamma_5)u_{\Lambda^+_c}}
{q^2_{2+3}-M^2_B-i\Gamma_{B}M_{B}},
\cr
i\mathcal{M}_{B'}^{(a)}&=&-\frac{ig_{KNB'}}
{M_KM_\pi}\frac{\overline{u}_p\gamma_5\Gamma_5
[\Delta_{\mu\nu}(q_{2+3})k^\mu_2k^\nu_1]
\Gamma_5(g^\mathrm{PV}_{\pi B'\Lambda_c}-g^\mathrm{PC}_{\pi B'\Lambda_c}\gamma_5)u_{\Lambda^+_c}}
{q^2_{2+3}-M^2_{B'}-i\Gamma_{B'}M_{B'}},
\cr
i\mathcal{M}^{(b)}_B&=&ig_{\pi NB}
\frac{\overline{u}_p\Gamma_5\Delta(q_{1+3})\Gamma_5\gamma_5
(g^\mathrm{PV}_{K B\Lambda_c}-g^\mathrm{PC}_{K B\Lambda_c}\gamma_5)u_{\Lambda^+_c}}
{q^2_{1+3}-M^2_B-i\Gamma_{B}M_{B}},
\cr
i\mathcal{M}_{B'}^{(b)}&=&-\frac{ig_{\pi NB'}}
{M_KM_\pi}\frac{\overline{u}_p\gamma_5\Gamma_5
[\Delta_{\mu\nu}(q_{1+3})k^\mu_1k^\nu_2]
\Gamma_5(g^\mathrm{PV}_{K B'\Lambda_c}-g^\mathrm{PC}_{K B'\Lambda_c}\gamma_5)u_{\Lambda^+_c}}
{q^2_{1+3}-M^2_{B'}-i\Gamma_{B'}M_{B'}},
\cr
i\mathcal{M}^{(c)}_V&=&ig_{\pi KV}\frac{\overline{u}_p
\left[\slashed{q}_{1-2}-\slashed{q}_{1+2}(M^2_\pi-M^2_K)/M^2_V
\right]
\gamma_5(g^\mathrm{PV}_{VN\Lambda_c}-g^\mathrm{PC}_{VN\Lambda_c}\gamma_5)
u_{\Lambda^+_c}}{M^2_{12}-M^2_V-i\Gamma_VM_V},
\end{eqnarray}
\end{widetext}
where the propagators for the spin-$1/2$ and spin-$3/2$ baryons are defined by
\begin{eqnarray}
\label{eq:PRO}
\Delta(q) = (\slashed{q} + M_B), &{}&\nonumber \\
\Delta_{\mu\nu}(q) = (\slashed{q} + M_B)&{}&\Bigl[g_{\mu\nu}
-\frac{1}{3} \gamma_\mu \gamma_\nu 
-\frac{1}{3M_B} (\gamma_\mu q_\nu - \gamma_\nu q_\mu) \nonumber \\
&{}&-\frac{2}{3M^2_B} q_\mu q_\nu \Bigr].
\end{eqnarray}

Now, the \textit{resonance-band patterns} on the Dalitz plot
are discussed in detail, 
as these patterns indicate nontrivial interferences between 
the resonance contributions and additional contributions. 
By carefully examining the $\Lambda(1670)$ band 
at $M^2_{K^-p}\approx2.79\,\mathrm{GeV}^2$ 
in the Dalitz plot given in Fig.~\ref{fig:fig1}(a), 
quite different patterns are observed between the left and right 
sides of the $K^*$ band. 
Moreover, the $K^*$ band exhibits a nontrivial pattern as well, i.e., 
it is distorted in the region where interference 
with the $\Lambda(1670)$ band occurs. 
To interpret this complicated pattern in 
the $\Lambda(1670)$-$K^*$ interference region, 
we consider the $\eta$-$\Lambda$ loop in a simple model; 
the loop channel opens at $(M_\eta+M_\Lambda)^2=2.767\,\mathrm{GeV}^2$, 
and it can cause complicated structures, 
such as a cusp~\cite{Bugg:2008wu}. 
The relevant Feynman diagram for describing the 
$\eta$-$\Lambda$ loop 
is depicted in Fig~\ref{fig:fig1}(d). This simple diagram is important for
the following reasons. 
First, all the vertex structures are theoretically known, 
i.e., the weak $\pi\Lambda\Lambda^+_c$ and 
strong $\eta\Lambda\Lambda$ vertices are given 
by HQEFT~\cite{Sharma:1998rd}  
and the Nijmegen soft-core potential~\cite{Rijken:2010zzb} 
as shown in Eq.~(\ref{eq:LAG}). 
The $\eta\overline{K}p\Lambda$ vertex is characterized by 
the Weinberg-Tomozawa (WT) chiral interaction:
\begin{eqnarray}
\mathcal{L}_{\Phi\Phi' BB'}=-\frac{ig_{\Phi\Phi' BB'}}{4f^2_\Phi}
\overline{B}(\Phi'\slashed{\partial}\Phi-\Phi\slashed{\partial}\Phi')B,
\label{eq:WT}
\end{eqnarray}
where $\Phi$ and $B$ denote the octet pseudo-scalar meson 
and baryon, respectively. 
Explicitly for the diagram given in Fig.~\ref{fig:fig1}(d), 
we use $g_{K^-p\eta\Lambda}=3/2$~\cite{Hyodo:2011ur}, 
$g_{\eta\Lambda\Lambda}=-6.86$~\cite{Hyodo:2011ur}, 
and $f_\Phi=f_\pi\times1.123\approx105$ MeV~\cite{Miyahara:2015cja} 
for numerical calculations. 
Second, as discussed in Ref.~\cite{Miyahara:2015cja}, 
the $I=1$ meson-baryon channel is suppressed similar to 
the $\eta$-$\Sigma^0$ one, 
in terms of the strong di-quark configuration inside $\Lambda^+_c$. 

Using the relevant interaction Lagrangians for the vertices 
in Eqs.~(\ref{eq:LAG}) and (\ref{eq:WT}), 
the diagram with a meson-baryon loop is computed as follows:
\begin{widetext}
\begin{eqnarray}
\label{eq:LOOP}
i\mathcal{M}_{\eta\Lambda}
&=&-\frac{g_{K^-p\eta\Lambda}\,g_{\eta\Lambda\Lambda}\theta_{\eta\Lambda}}{4f^2_\Phi}
\int\frac{d^4q}{(2\pi)^4}\left[
\frac{\overline{u}_p(\slashed{q}+\slashed{k}_3)(\slashed{q}_{2+3}-\slashed{q}+M_\Lambda)
\gamma_5(\slashed{q}_{2+3}+M_\Lambda)(g^\mathrm{PV}_{\pi^+ \Lambda\Lambda^+_c }
-\gamma_5g^\mathrm{PC}_{\pi^+\Lambda\Lambda^+_c })u_{\Lambda^+_c}}
{[q^2-M^2_\eta][(q-q_{2+3})^2-M^2_\Lambda][q_{2+3}^2-M^2_\Lambda]}\right]
\cr
&\approx&\frac{ig_{K^-p\eta\Lambda}\,g_{\eta\Lambda\Lambda}\theta_{\eta\Lambda}}{4f^2_\Phi}
\,G(q^2_{2+3})\,
\frac{\overline{u}_p(\slashed{q}_{2+3}+\slashed{k}_3-M_\Lambda)\gamma_5(\slashed{q}_{2+3}+M_\Lambda)
(g^\mathrm{PV}_{\pi^+ \Lambda\Lambda^+_c }
-\gamma_5g^\mathrm{PC}_{\pi^+\Lambda\Lambda^+_c })u_{\Lambda^+_c}}{[q_{2+3}^2-M^2_\Lambda]},
\end{eqnarray}
\end{widetext}
where $q_{i+j}\equiv k_i+k_j$, The step function for 
the $\eta$-$\Lambda$ channel threshold 
is defined by $\theta_{\eta\Lambda}=\theta(M_{\overline{K}p}-M_\eta-M_\Lambda)$. 
In deriving the $\eta$-$\Lambda$ loop integral in Eq.~(\ref{eq:LOOP}), 
the on-shell factorization~\cite{Hyodo:2011ur} is employed, 
which assumes that $\Lambda$ in the loop is almost its on-shell. 
This approximately satisfies the following relationship: 
\begin{equation}
\label{eq:ONFAC}
(\slashed{q}_{2+3}-\slashed{q})u_\Lambda\approx M_\Lambda u_\Lambda\to
\slashed{q}\approx (\slashed{q}_{2+3}-M_\Lambda). 
\end{equation}
The loop divergence is regularized by the dimensional regularization 
and the meson-baryon propagating function $G$ is given by~\cite{Nam:2003ch} 
\begin{widetext}
\begin{eqnarray}
\label{eq:GFUN}
G_\mathrm{dim}(q^2_{2+3})&=&i\int\frac{d^4q}{(2\pi)^4}
\frac{2M_\Lambda}{[q^2-M^2_\eta][(q-q_{2+3})^2-M^2_\Lambda]}
\cr
&=&
\frac{2M_\Lambda}{16\pi^2}\left[
\frac{M^2_\eta-M^2_\Lambda+q^2_{2+3}}{2q^2_{2+3}}\ln\frac{M^2_\eta}{M^2_\Lambda}
+\frac{\xi}{2q^2_{2+3}}\ln\frac{M^2_\eta+M^2_\Lambda-q^2_{2+3}-\xi}
{M^2_\eta+M^2_\Lambda-q^2_{2+3}+\xi}
\right]+\frac{2M_\Lambda}{16\pi^2}\ln\frac{M^2_\Lambda}{\mu^2},
\end{eqnarray}
\end{widetext}
where $\xi$ is defined by
\begin{equation}
\label{eq:}
\xi\equiv\sqrt{[q^2_{2+3}-(M^2_\eta-M^2_\Lambda)^2][q^2_{2+3}-(M^2_\eta+M^2_\Lambda)^2]}.
\end{equation}
The subtraction scale for the regularization is chosen as $\mu=630$ MeV, 
which is responsible for dynamically reproducing 
the $S=-1$ hyperon resonances 
in the couple-channel chiral-unitary model (ChUM)~\cite{Hyodo:2011ur}.

The total amplitude consists of the relevant contributions as follows:
\begin{eqnarray}
\label{eq:AMP}
i\mathcal{M}_\mathrm{total}&=&
i\sum_{\Lambda^*}c_{\Lambda^*}\mathcal{M}_{\Lambda^*}F_{\Lambda^*}
+ic_{\Lambda}\mathcal{M}_{\Lambda}F_{\Lambda}
+ic_{\Sigma}\mathcal{M}_{\Sigma}F_{\Sigma} \nonumber \\
&+&ic_{N}\mathcal{M}_{N}F_{N}
+ic_{\Delta}\mathcal{M}_{\Delta}F_{\Delta}
+ic_{K^*}\mathcal{M}_{K^*}F_{K^*} \nonumber \\
&+&ic_{\eta{\textrm -}\Lambda}\mathcal{M}_{\eta{\textrm -}\Lambda}F_{\eta{\textrm -}\Lambda}
,
\end{eqnarray}
where the coefficients $c_h$ and $F_h$ denote the relative phase factor 
and phenomenological form factor for the hadron $h$, respectively. 
Note that the $\Lambda^*$ ($N$) contribution 
is only given in the charged (neutral) channel. 
The phenomenological form factors are considered 
because the hadrons are spatially extended objects. In the present work, 
we employ the following parameterization:
\begin{equation}
\label{eq:FF}
F_h(q^2_h)=\frac{\mathcal{C}\Lambda^4}{\Lambda^4+(q^2_h-M^2_h)},
\end{equation}
where $q_h$ and $M_h$ denote the momentum transfer and mass of 
the intermediate hadron, respectively. 
For brevity, we fix the cutoff mass $\Lambda=1.0$ GeV 
for all hadrons throughout the present work. 
To compensate for this simplification regarding the cutoff choices, 
we introduce a phenomenological parameter 
$\mathcal{C}$ in Eq.~(\ref{eq:FF}), which will be adjusted to reproduce 
the decay width of the experimental data.

\section{Numerical results and Discussions}

In this Section, we discuss the numerical results for the 
$\Lambda_c^+$ decays. 
We first show the numerical results for  
the Dalitz plots for the charged ($\Lambda^+_c\to\pi^+K^-p$) and 
neutral ($\Lambda^+_c\to\pi^0\overline{K}^0p$) channels 
in Fig.~\ref{fig:fig2}. 
According to the calculated Dalitz plot density, 
simulated events are generated over the phase
space available. We assume a uniform experimental acceptance
for the $\pi\overline{K}p$ phase space.
Note that all unknown weak coupling constants 
and phase factors in Eq.~(\ref{eq:AMP}) are carefully determined 
to reproduce the data, 
focusing on the complicated interference patterns in the Dalitz plot 
of the Belle data~\cite{Yang:2015ytm}, as listed in Table~\ref{tab:tab1} 
and \ref{tab:tab2}. 
To determine the phenomenological parameter 
for the form factor in Eq.~(\ref{eq:FF}), 
which will provide the overall strength of the decay width, 
we employ the experimental data for the decay ratio between 
the charged and neutral channels. 
Considering that the $\Lambda^+_c(2286,1/2^+)$ baryon has a life time of
$\tau_{\Lambda^+_c}=(2\times10^{-13})$ s, which corresponds to 
$\Gamma_{\Lambda^+_c}=(3.29\times10^{-9})$ MeV~\cite{Tanabashi:2019oca} 
and employing the PDG values of the partial decay ratios 
for the charge and neutral decays, 
the resulting relationship is 
\begin{eqnarray}
\label{eq:decay}
\Gamma_{\Lambda^+_c\to \pi^+K^-p}&=&(2.09 \times 10^{-10})\,\mathrm{MeV},~~~~~~\nonumber \\
\Gamma_{\Lambda^+_c\to \pi^0\overline{K}^0p}&=&(1.31 \times 10^{-10})\,\mathrm{MeV}.
\end{eqnarray}
We note that $\overline{K}^0$ is a mixture of $K^0_S$ and $K^0_L$ 
in the same proportion, if we ignore the CP violation. 
Furthermore, in general, in the experimental data, 
the $K^0_S$ was measured for the $\Lambda^+_c$ decay, as shown in 
\cite{Tanabashi:2019oca}. 
Thus, we simply doubled the experimental partial-decay ratio 
in our theoretical calculations, as shown in Eq.~({\ref{eq:decay}). 
To reproduce these decay widths, 
we choose $\mathcal{C}_\mathrm{charged}=5.25$ 
and $\mathcal{C}_\mathrm{neutral}=3.15$ in Eq.~(\ref{eq:FF}).

\begin{center}
\begin{table}[h]
\begin{tabular}{c|c||c|c}\hline
 & $c_h$ & & $c_h$ \\ \hline
$p(939)$  & $-0.9i$ & $\Delta(1232)$  &  1  \\  
$\Lambda(1116)$  & $-0.9i$ & $\Sigma(1193)$ & $-0.9i$\\
$\Lambda^*(1520)$ & 1 & $K^*(892)$ & $i$ \\
$\Lambda^*(1670)$ & 1 & $\eta$-$\Lambda$ & $1.8i$\\
$\Lambda^*(1890)$ & 1 & & \\
\hline
\end{tabular}
\caption{Relative phase factors for the amplitudes in Eq.~(\ref{eq:AMP}). 
Note that the phase factor for the $\eta$-$\Lambda$ loop 
contribution is also introduced. }
\label{tab:tab2}
\end{table}
\end{center}

\begin{figure}[t]
\includegraphics[width=0.48\textwidth]{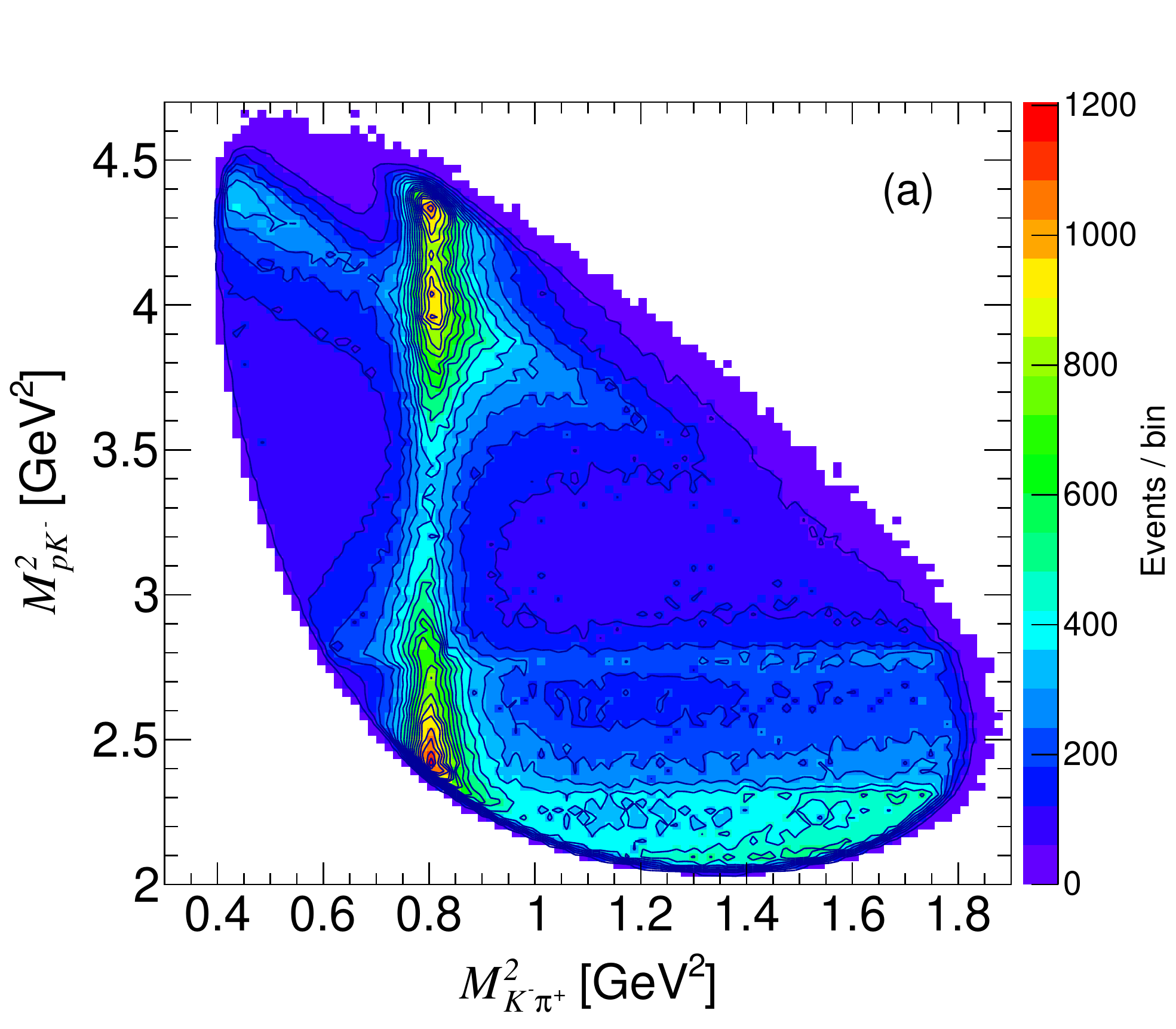}
\includegraphics[width=0.48\textwidth]{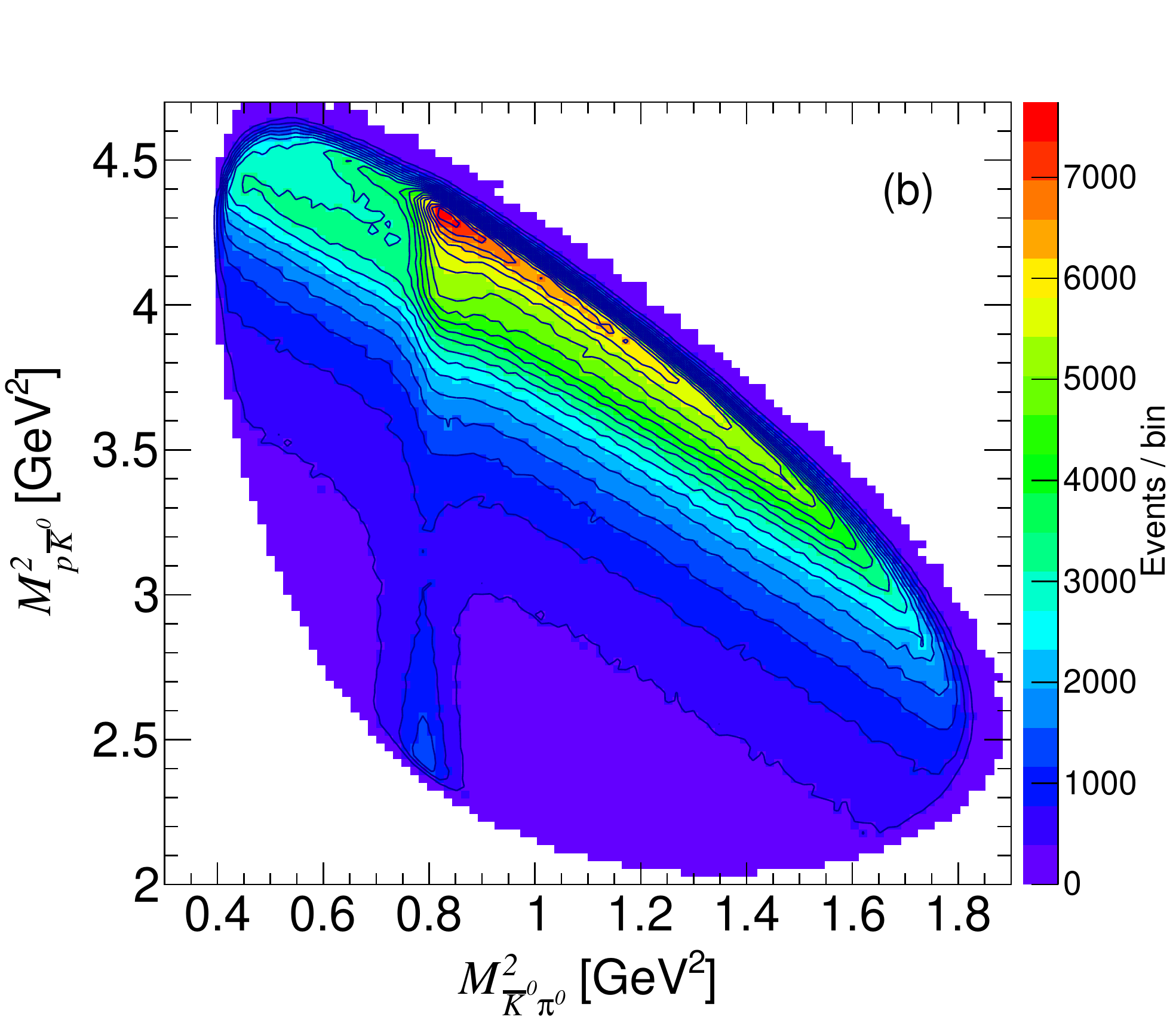}
\caption{(Color online) Dalitz plots for (a) 
$\Lambda^+_c\to \pi^+K^-p$ for $\Lambda^+_c\to \pi^+K^-p$ 
and (b) $\Lambda^+_c\to \pi^0\overline{K}^0p$.}       
\label{fig:fig2}
\end{figure}

As clearly shown in Fig \ref{fig:fig2}(a), the contributions from 
$\Lambda(1520)$ and $\Lambda(1670)$, as the horizontal bands, are 
consistent with the data. 
We also note that the small contribution from $\Lambda(1890)$ 
increases the strength of the $K^*$ band in the interference region. 
The $\Delta(1232)$ contribution provides the sloped band 
in the upper region of the Dalitz plots. 
The nontrivial pattern shown in the $K^*$-$\Lambda(1670)$ 
interference region is qualitatively reproduced by the 
$\eta$-$\Lambda$ loop. 
We verified that the $\Lambda(1670)$ band becomes smooth 
and shows no complicated interference patterns without 
the $\eta$-$\Lambda$ loop. 
From this observation, we conclude that the nontrivial interference 
patterns in the region of interest are due to the $\eta$-$\Lambda$ loop 
and nontrivial phase factors given in Table~\ref{tab:tab2}. 
Moreover, this observation indicates that the coherent sum 
using Breit-Wigner amplitudes does not accurately represent reality. 

Once all the couplings are considered, including the phase factors 
shown in Tables~\ref{tab:tab1} and \ref{tab:tab2}, 
we attempt to predict the neutral channel, i.e., 
$\Lambda^+_c\to\pi^0\overline{K}^0p$, 
which has not yet been reported experimentally. 
Note that we choose the same phase factor $-0.9i$ for the proton as 
for the $\Lambda$ and $\Sigma$ BKGs. 
The numerical result for the Dalitz plot is depicted 
in Fig.~\ref{fig:fig2}(b). 
Because there are no $I=0$ hyperon resonances in this channel, 
we observe dominant contributions from the $K(892)$, $\Delta(1232)$, 
and $p$, in addition to the small ground-state $\Sigma$ background. 
Note the lack of an $I=0$ channel opening effect here. 

\begin{figure}[h!]
\stackinset{r}{1.cm}{t}{0.5cm}{(a)}{
\includegraphics[width=0.48\textwidth]{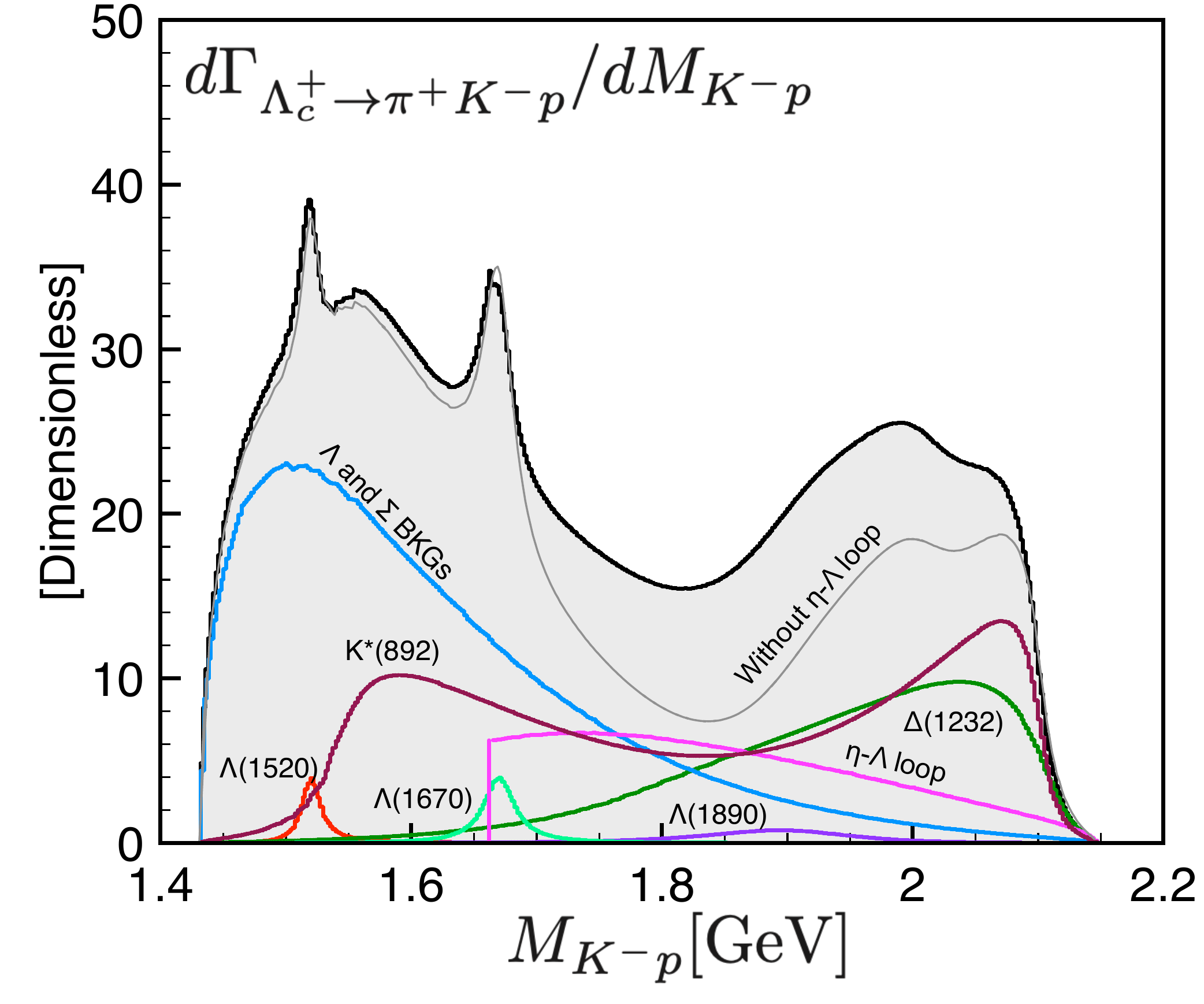}
}
\stackinset{r}{1.cm}{t}{0.5cm}{(b)}{
\includegraphics[width=0.48\textwidth]{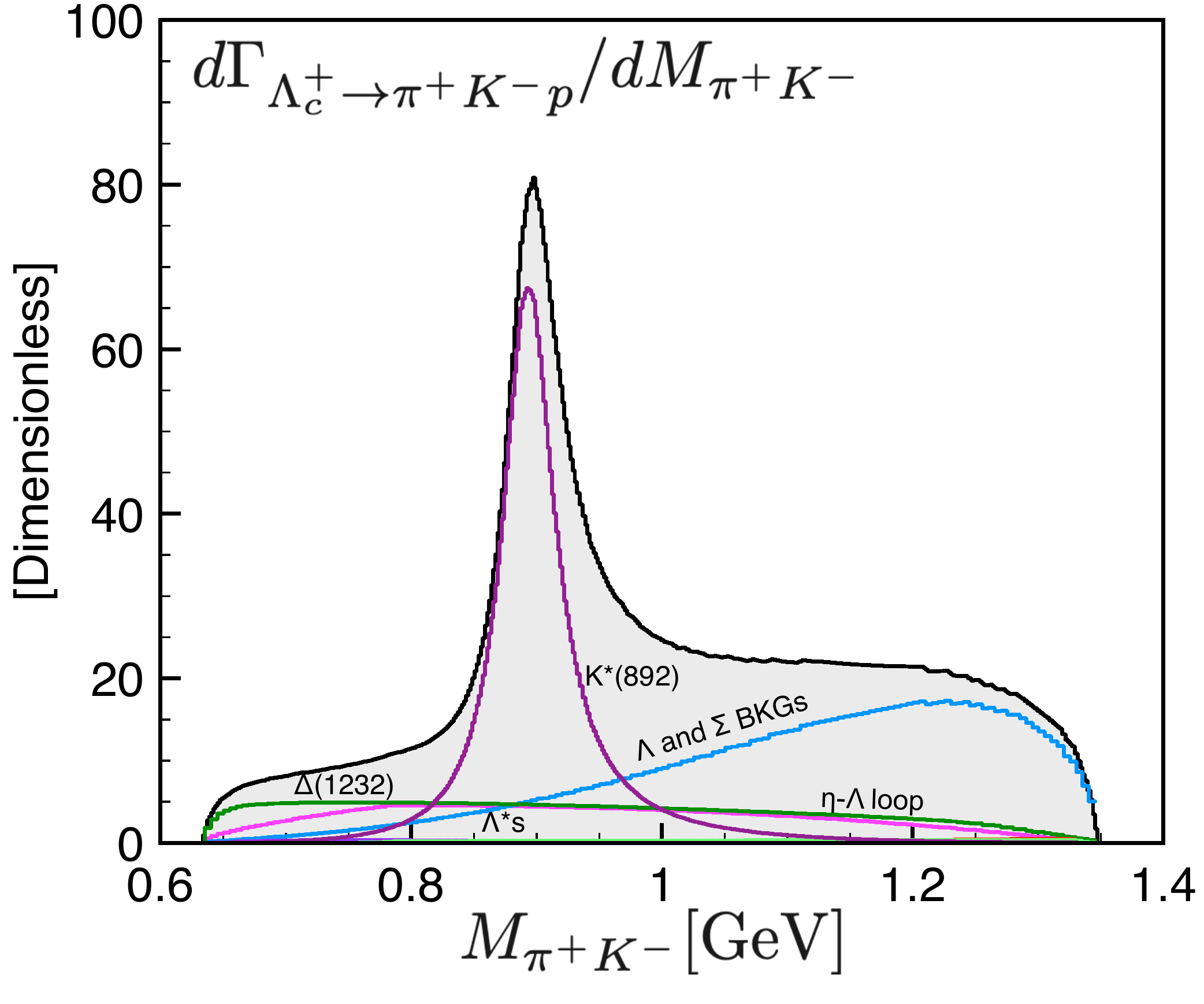}
}
\caption{(Color online) Invariant-mass distributions for the charged channel 
as functions of (a) $M_{K^-p}$ and (b) $M_{\pi^+K^-}$.}          
\label{fig:fig3}
\end{figure}
\begin{figure}[h!]
\stackinset{r}{1.cm}{t}{0.5cm}{(a)}{
\includegraphics[width=0.48\textwidth]{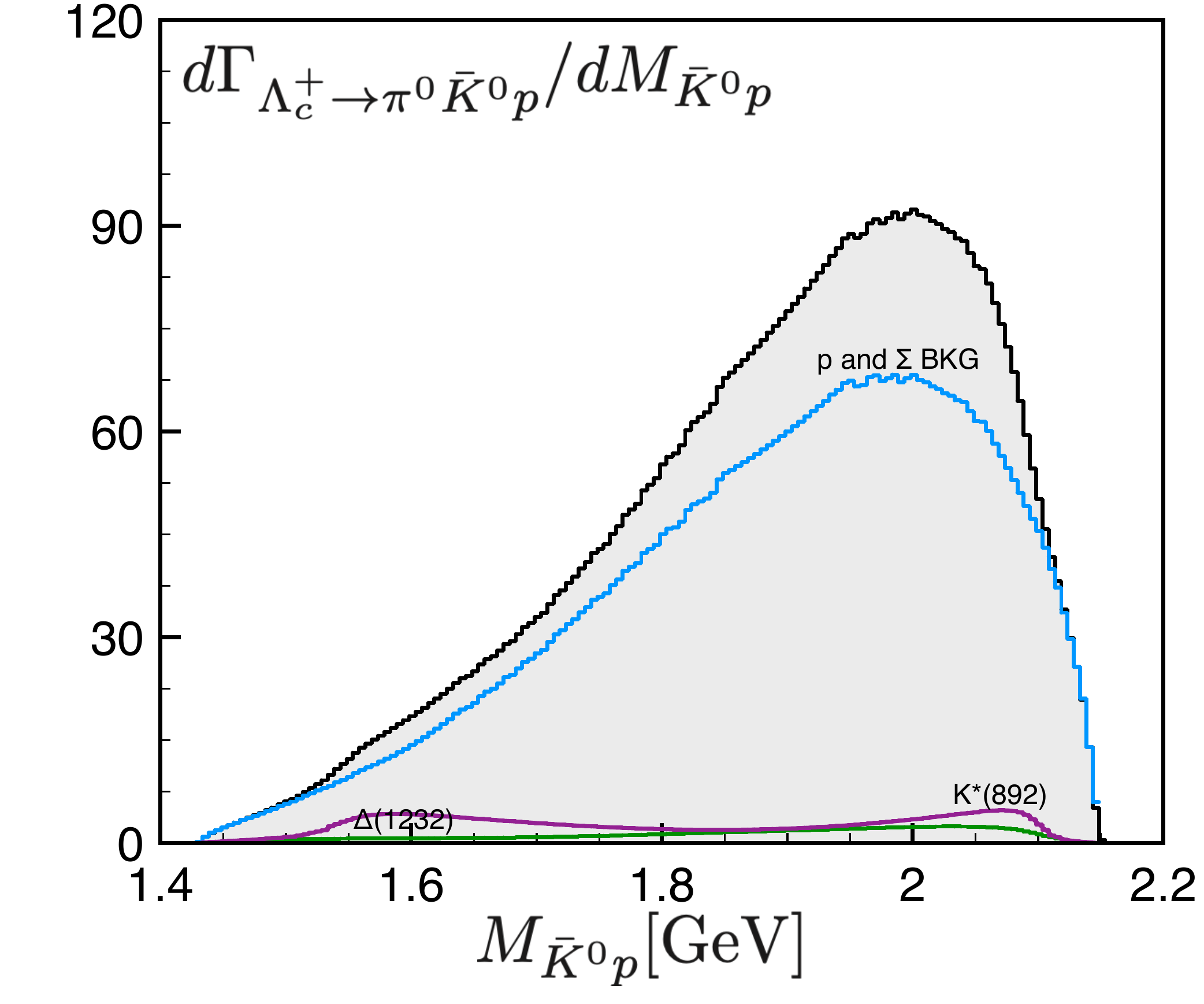}
}
\stackinset{r}{1.cm}{t}{0.5cm}{(b)}{
\includegraphics[width=0.48\textwidth]{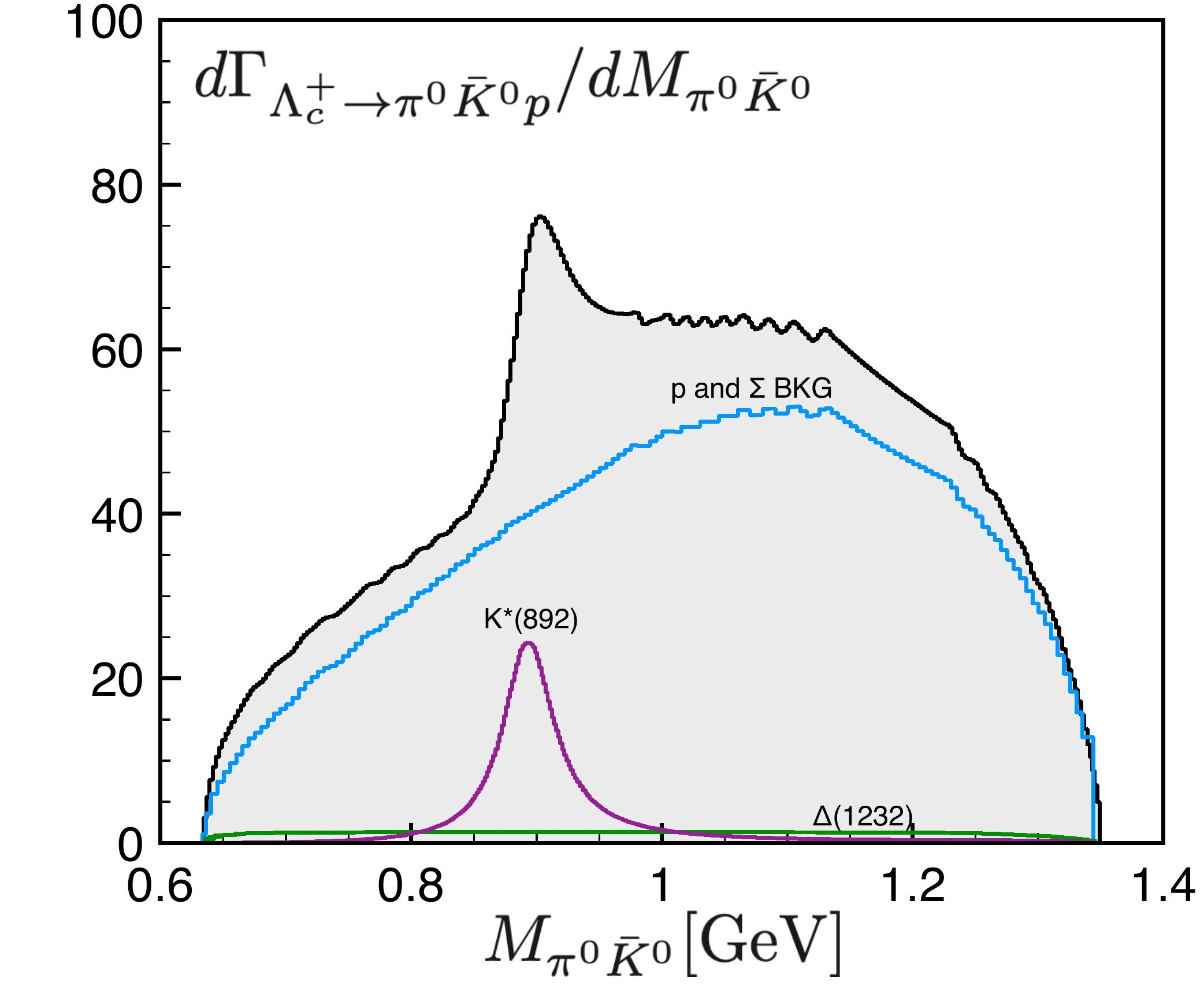}
}
\caption{(Color online) Invariant-mass distributions for the neutral channel 
as functions of (a) $M_{\overline{K}^0p}$ and (b)$M_{\pi^0\overline{K}^0}$.}          
\label{fig:fig4}
\end{figure}

With these data, it is possible to investigate 
the invariant-mass distributions. 
In Figs.~\ref{fig:fig3}(a) and (b), we draw their numerical results
as functions of $M_{K^-p}$ and $M_{\pi^+K^-}$, respectively, 
for the charged channel. 
The curves for each contribution are also shown to highlight 
their complicated combination for the distribution. 
As shown in Fig.~\ref{fig:fig3}(a), peaks for the $\Lambda(1520)$ 
and $\Lambda(1670)$ contributions are easily observed. 
The two bumps are generated by the $K(892)$ contribution 
at $M_{K^-p}\approx1.6$ GeV and $2.1$ GeV, whereas the ground-state 
$\Lambda$ and $\Sigma$ BKGs dominate the low-invariant-mass region. 
Interestingly, because of the interference between 
the $\eta$-$\Lambda$ loop and $\Lambda(1670)$, 
a peak-like sharp structure appears in the vicinity of 
$M_{K^-p}\approx1.67$ GeV. 
Notably, this peak-like structure is inevitable when reproducing 
the nontrivial interference pattern observed 
in the charged-channel Dalitz plot from the experimental data. 
In Fig.~\ref{fig:fig3}(b), the distribution as a function of 
$M_{\pi^+K^-}$,
the $K(892)$ dominates in addition to 
the considerable contributions from the hyperon BKGs and $\Delta$, 
while the $\eta$-$\Lambda$ loop contribution is not obvious.
 
Similarly, Figs.~\ref{fig:fig4}(a) and (b) provide 
the numerical results for the invariant-mass distributions 
for the neutral channel, i.e., $\Lambda^+_c\to\pi^0\overline{K}^0p$, 
which has not yet been reported experimentally. 
As shown in Fig.~\ref{fig:fig4}(a), considering 
the $I=1$ channel suppression, DCS~\cite{Miyahara:2015cja}, 
and absence of $I=0$ hyperons, 
the distribution as a function of $M_{\overline{K}^0p}$ does not 
show peak-like structures at all.  
Instead, the proton-pole contribution dominates the distribution 
and provides a large bump structure at $M_{\overline{K}^0p}\approx2$ GeV, 
whereas the ground-state $\Sigma$, $\Delta(1232)$, 
and $K(892)$ contributions are small. 
In Fig.~\ref{fig:fig4}(b), we show the distribution 
as a function of $M_{\pi^0\overline{K}^0}$. 
We observe a clear peak generated from the $K(892)$ contribution 
on top of the $p$-pole contribution. 

Finally, an important aspect of the results is 
the \textit{considerably} narrow band structure 
at $M_{K^-p}\approx1.67$ GeV, as shown in the Dalitz plot 
in Fig.~\ref{fig:fig4}(a). 
Compared with the width of $\Lambda(1520)$, the band width is
similar to that of $\Gamma_{\Lambda(1520)}\approx 15$ MeV. 
There are several possibilities 
to explain the narrow band. First, there have thus far been no such 
hyperon resonances with such a narrow width at $M_{Y^*}=(1.6\sim1.7)$ 
GeV~\cite{Tanabashi:2019oca}. 
Given this observation, the band might signal a missing hyperon 
resonance, as reported in Ref.~\cite{Kamano:2015hxa}, 
i.e., $\Lambda(1671)(J^p=3/2^+)$ with a width of $\Gamma\approx10$ MeV. 
However, we verified that it is difficult to reproduce 
the \textit{nontrivial} interference pattern 
by combining the two Breit-Wigner-type amplitudes, i.e., 
the new hyperon resonance and $K(892)$ contributions.
Second, there are 
complicated interferences between the known hyperon resonances 
at $M_{\overline{K}p}=(1.6\sim1.7)$ GeV, such as $\Lambda(1600)$, 
$\Lambda(1670)$, $\Lambda(1690)$.
However, we verified that it is almost impossible to form 
this peak structure numerically. 
Third, the interference between the $\Lambda(1670)$ and  
$\eta$-$\Lambda$ loop channels has been explored in this work 
and the \textit{nontrivial} interference pattern 
and narrow peak-like structure have been qualitatively explained. 
However, as discussed in Section II, theoretical considerations simplify 
the interpretation; for example, the strong di-quark correlation 
inside $\Lambda^+_c$ and color-factor 
for the non-strange resonances provide constraints.
 
\begin{figure}[h!]
\includegraphics[width=0.48\textwidth]{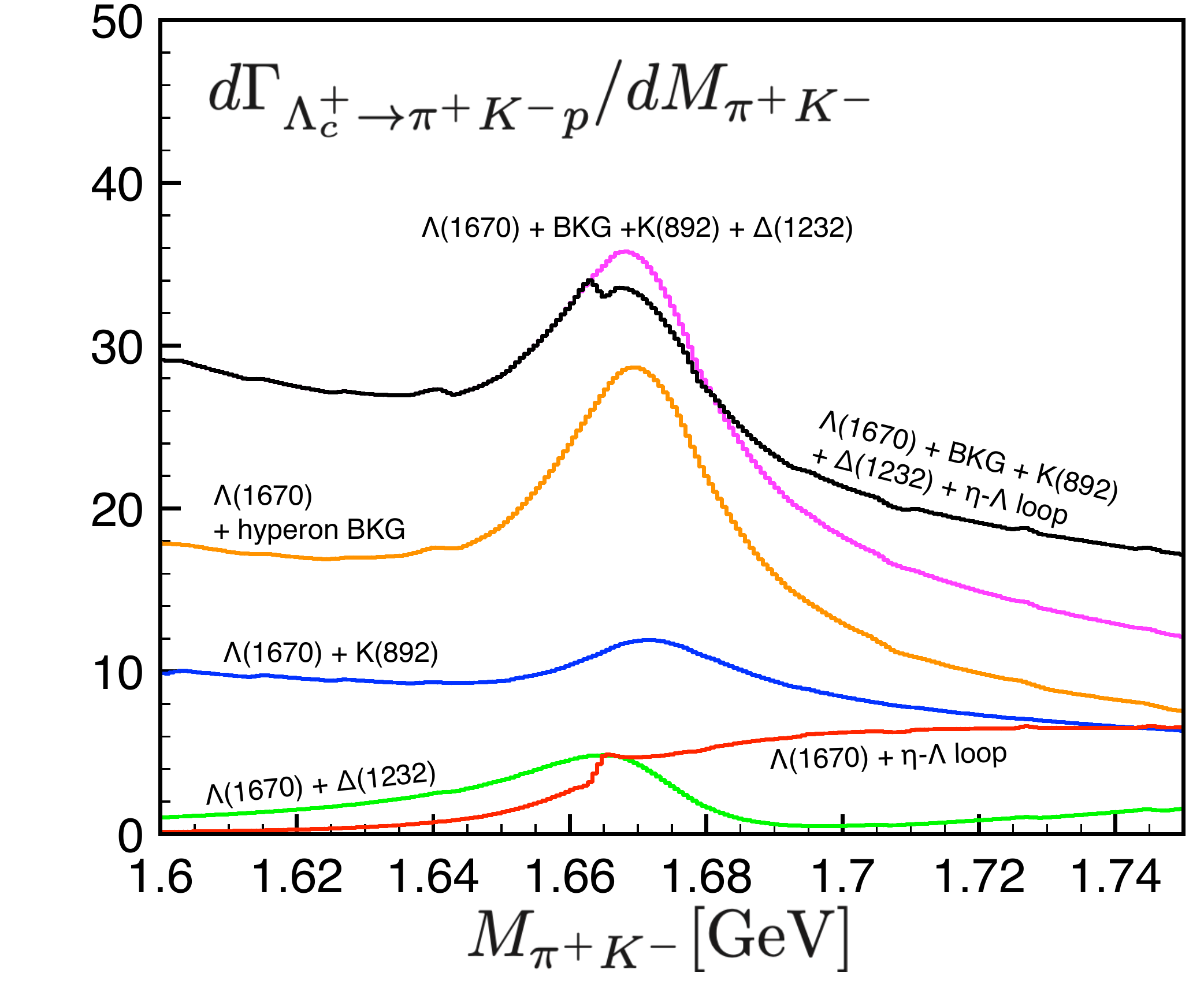}
\caption{(Color online) The interference patterns 
between $\Lambda(1670)$ and other contributions. 
See the text for details.}       
\label{fig:fig5}
\end{figure}

The interference effects between the $\Lambda(1670)$ 
and other channels are illustrated in the vicinity of 
$M_{K^-p}\approx 1.67$ GeV in Fig. \ref{fig:fig5}.
It is obvious that the $\Delta(1232)$, $K(892)$, and hyperon BKG
channels enhance the $\Lambda(1670)$ structure and result in a large
peak at 1.67 GeV. On the other hand, 
A strong destructive interference effect
between the $\Lambda(1670)$ and $\eta$-$\Lambda$ loop channels is 
observed and provides the sharp peak-like structure 
near the $\eta$-$\Lambda$ threshold.

\section{Summary}

In this study, we investigated the excited hyperon production 
in the $\Lambda^+_c\to\pi^+K^-p$ (charged) and 
$\Lambda^+_c\to\pi^0\overline{K}^0p$ (neutral) decays
within the effective Lagrangian approach. 
We determined relevant model parameters for phenomenological 
form factors at the tree-level Born approximation based on
experimental data for $\Lambda^+_c\to\pi^+K^-p$~\cite{Yang:2015ytm} 
and the known decay branching ratios for 
$\Lambda^+_c$ ~\cite{Tanabashi:2019oca}. 
We list important observations as follows:
\begin{itemize}  
\item To reduce the number of hyperons considered, we accounted for 
the strong di-quark correlation inside $\Lambda^+_c$ and color-factor constraint of the quarks from $W^+$, 
which makes it possible 
to drop the $\Sigma^*$, $\Delta^*$, and $N^*$ contributions 
in the numerical calculations, although there seem non-negligible 
$I=0$ contributions in reality. As a result of these assumptions, 
the theoretical uncertainties were substantially diminished. 
\item Regarding the charged-channel decay, 
on top of the hyperon backgrounds, 
$\Lambda(1520)$ and $\Lambda(1670)$ exhibited obvious peaks 
in the invariant-mass distribution in addition to the bump structure, 
caused by the interference between the $\Delta(1232)$ and $K(892)$. 
The Belle experimental data for the Dalitz plot are qualitatively 
reproduced.
\item A nontrivial interference pattern was observed 
in the charged-channel Dalitz plot at the interference region 
between $\Lambda(1670)$ and $K(892)$, 
which can be explained successfully by including the $\eta$-$\Lambda$ 
loop contribution.
Moreover, this complicated interference generates 
the peak-like structure at 
$M_{K^-p}\approx1.655$ MeV. 
\item Given the previous point, we did not observe a sharp peak 
(band) structure in the neutral channel, 
as the $I=0$ meson-baryon channel opening was absent. 
By contrast, a sharp peak-like structure, together with a nontrivial interference pattern in experiments,  
indicate a small but finite $I=1$ channel opening is possible 
as a next-to-leading contribution 
to the di-quark configuration of $\Lambda^+_c$. 
\item Regarding the neutral channel with the model parameters, determined by the charged-channel data and theory models, 
a strong background from the proton-pole contribution was observed.
The result was an absence of clear peak-like structures 
in the invariant-mass distribution 
as a function of $M_{\overline{K}^0p}$, 
while the $K(892)$ contribution showed an obvious peak 
on top of the proton background in the invariant-mass distribution 
as a function of $M_{\pi^0\overline{K}^0}$.
\end{itemize}  

As discussed, the charmed-baryon decays are good places to
study weak interactions in the context of Cabibbo-favored decays, 
isospin selections, and confinement 
corresponding to the color factor of $W^+$ decays. 
Moreover, the internal structure of the involved hadrons 
can be probed by comparing the di-quark scenario with experiments. 
The channel-opening effects are clearly also important. Therefore, 
the charmed-baryon decay with $S=0$ hadrons is an interesting 
alternative to understanding weak and strong interactions. 
Related studies are in progress and will appear elsewhere.

\acknowledgments
The authors are grateful for the fruitful discussions with K.~S.~Choi (KAU). 
This work was partially supported by the National Research Foundation 
(NRF) of Korea 
(No.2018R1A5A1025563, No.2017R1A2B2011334, No. 2018R1A6A3A01012138). 
The work of S.i.N. was also supported in part by 
an NRF grant (No.2019R1A2C1005697).

%

\end{document}